\documentclass[twocolumn]{revtex4-2}
\usepackage{amsmath, amssymb}
\usepackage{graphicx}

\begin{document}

\title{Einstein's perihelion formula and its generalization}
\author{Maurizio M. D'Eliseo}
\affiliation{Osservatorio S.Elmo - Via A.Caccavello 22, 80129 Napoli, Italia}
\date{\today}

\begin{abstract}
Einstein's perihelion advance formula can be given a geometric interpretation in terms of the curvature of the ellipse. The formula can be obtained by splitting the constant term of an auxiliary polar equation for an elliptical orbit into two parts that, when combined, lead to the expression of this relativistic effect. Using this idea, we develop a general method for dealing with orbital precession in the presence of central perturbing forces, and apply the method to the determination of the total (relativistic plus Newtonian) secular perihelion advance of the planet Mercury.
\end{abstract}

\maketitle

\section{Introduction}
A classic calculation in the scientific literature is the derivation of the formula by which Einstein explained an apparent anomaly of the observed motion of the planet Mercury. A planet's perihelion remains fixed under a pure inverse-square gravitational force, so any shift indicates, as first realized by Newton, either a different force law or the presence of perturbing forces. The perihelion of Mercury is observed to precess---after correction for known planetary perturbations---at the rate of about 43 seconds of arc per century, and this residue is exactly predicted by the theory of general relativity.

To approximately derive the relativistic contribution to the precession (there are further corrections of negligible relevance), it is not necessary to completely solve the relativistic orbit equation. In his original derivation, Einstein came upon an elliptic integral, which he managed to compute approximately \cite{einstein1915}. Since then, a host of authors have taken alternate approaches to illuminate various aspects of this problem \cite{deliseo2011, davies1983, gauthier1987, garavaglia1987, farina1987, stump1988, mcdonald1990, doggett1991, cornbleet1993, dean1999, brill1999, deliseo2007a, lemmon2009}. 
Our approach to the subject arises from the interplay of two quite different methodological strategies, which we can define as the local and the global. The first shows that the perihelion precession produced by a perturbing force can be traced back to a steady action along the entire orbit. The second deals directly with this secular effect by splitting the constant term of an auxiliary polar equation of an elliptical orbit into two parts according to a specific criterion, without needing to know what takes place during the motion. Our method is applicable to a broad class of perturbing central forces and provides the leading term of the secular perihelion shift.

\section{Relativistic Equation}
The polar equation for the orbit of a planet, considered as a test particle subject to a central force $f(u)$, is given by the well-known expression:
\begin{align}
u'' + u = \frac{f(u)}{l^2 u^2}.
\end{align}
Here, $u(\theta) = 1/r$, with $r$ the distance from the origin, while $l$ is the (constant) magnitude of the orbital angular momentum, and a prime denotes differentiation with respect to the independent variable, which in this case is the angle $\theta$.

In general relativity, the spherically symmetric Schwarzschild solution to Einstein’s field equation corresponds, for a weak field, to a function $f(u)$ consisting of two attractive parts, the classical inverse-square force and a small corrective term \cite{dinverno2001}:
\begin{align}
f(u) = \lambda u^2 + 3\alpha l^2 u^4.
\end{align}
Here, $\lambda = GM$, with $G$ the universal gravitational constant and $M$ the mass of the star, and $\alpha = \lambda / c^2$, with $c$ the speed of light. The parameter $\alpha$ has the dimension of a length and is called the gravitational radius; for the Sun we have $\alpha \approx 1.477 \, \text{km}$, a very small value compared to typical orbital radii in our solar system. From Eqs.~(1) and (2), we get the relativistic orbit equation:
\begin{align}
u'' + u = \frac{\lambda}{l^2} + 3\alpha u^2,
\end{align}
which represents an oscillator with a weak quadratic nonlinearity. This equation cannot be solved exactly. Using methods of perturbation theory, a bounded periodic (planetary case) approximate solution can be painstakingly assembled to arbitrarily high order in the coupling constant $\alpha$, allowing a determination of the precession.

Our plan here is to bypass the solution of Eq.~(3) and, more generally, to extract directly from a perturbed orbital equation the leading precession term through a simple linearization process that consists of replacing the nonlinear term by a constant. This procedure will be discussed in detail after we have dealt with some basic aspects of the elliptical orbit.

\section{Elliptical Orbit}
The unperturbed orbit equation is \cite{grossmann1996}:
\begin{align}
u'' + u = \frac{\lambda}{l^2}.
\end{align}
The constant term $\lambda/l^2$ in this equation can be given a geometric meaning by exploiting a result of Newton’s that dates back to 1671. Newton found that a generic plane curve $u(\theta)$ satisfies the equation:
\begin{align}
u'' + u = \frac{1}{q \sin^3 \beta},
\end{align}
where $q$ is the radius of curvature and $\beta$ is the angle between the radial and tangential directions at any point on the curve.

Setting $\beta = \pi/2$, it follows that the orbit equation can be written in the form:
\begin{align}
u'' + u = \kappa,
\end{align}
where $\kappa \equiv 1/q = \lambda/l^2$ is the curvature at those particular points. For a planetary orbit this identification occurs in two circumstances: when the orbit is circular (in which case $\kappa = 1/r$ is always true) and at the extrema of an elliptical orbit (perihelion and aphelion). In the latter case, the curvature is maximal and can be simply expressed in terms of the ellipse parameters (eccentricity $e$, semi-major axis $a$, and semi-minor axis $b$).

The solution to Eq.~(6), written so as to highlight this geometrical aspect, is obtained by starting with the particular solution $u = \kappa$ and adding to it the periodic solution to the homogeneous equation. The orbit can thus be written in the form:
\begin{align}
u(\theta) = \kappa + \kappa e \cos(\theta - \chi),
\end{align}
where $e$ and $\chi$ are the two arbitrary constants that parametrize the family of solutions. This is the polar equation of an ellipse when the eccentricity $e$ lies in the open interval $(0, 1)$; the polar axis has been chosen so that $u$ attains its maximum value at $\theta = \chi$, the phase that identifies the position of the perihelion. The period of the solution is $2\pi$, so that the perihelia are located at $\theta = \chi + 2n\pi$, with $n = 0, 1, \dots$.

Now, the particular solution $u = \kappa$ of Eq.~(4) represents a circular orbit of radius $r = q$, and from Eq.~(7) we find that $\kappa = u(\chi + \pi/2)$. Then, by comparison with the standard polar equation for an ellipse, we deduce that the curvature is given by:
\begin{align}
\kappa = \frac{1}{q} = \frac{1}{a(1 - e^2)} = \frac{a}{b^2} \quad \text{at perihelion}.
\end{align}
\section{Angular Period Operator}
A definite integral provides global information about the behavior of a function in the interval of integration. As a limit of Riemann sums, it is the result of a pointwise accumulation process, and this proves useful for the detection of secular effects on the orbit. Unlike Einstein's elliptic integral, which is quite complicated to handle, the integral we shall use is elementary, because the integrand is the function $u(\theta)$, the sum of a constant and a cosine.

When plotted in rectangular coordinates $(\theta, u)$, with $\theta$ in the range $[\chi, \chi + 2\pi]$, the function $u = \kappa + \kappa e \cos(\theta - \chi)$ traces out a sinusoid of amplitude $\kappa e$ around the segment $u = \kappa$ of length $2\pi$. The oscillation begins and terminates at two successive perihelia, the points $(\chi, \kappa + \kappa e)$ and $(\chi + 2\pi, \kappa + \kappa e)$, while in between we have one aphelion at point $(\chi + \pi, \kappa - \kappa e)$. It follows that $\kappa$ is the average value of the elliptical solution over the interval $[\chi, \chi + 2\pi]$:
\begin{align}
\frac{1}{2\pi} \int_{\chi}^{\chi + 2\pi} u \, d\theta = \kappa,
\end{align}
and so $u = \kappa$ is the circular orbit of an imaginary planet associated with the planet that follows the elliptical orbit (7), with whom it shares the same period. The integral:
\begin{align}
\int_{\chi}^{\chi + 2\pi} u \, d\theta = 2\pi \kappa,
\end{align}
measures the area under the graph of $u$ in the interval $[\chi, \chi + 2\pi]$, and is equal to the area of the rectangle of width $2\pi$ and height $\kappa$. Although representing the area of a plane figure, this result can also be interpreted as the circumference of a circle of radius $\kappa$. Therefore, a division by $\kappa$ will provide the period of the orbit, i.e., the distance (from $\chi$) along the $\theta$-axis after which the function $u$ repeats itself.

In general, given a real number $s$, we can consider the integration:
\begin{align}
\hat{P} u(s\theta) \equiv \frac{1}{\kappa} \int_{\chi}^{\chi + 2\pi/s} u(s\theta) \, d\theta = 2\pi \frac{1}{s},
\end{align}
as the action of an operator $\hat{P}$ that, when acting on the function $u(s\theta)$, results in the angular period (defined as the angle separating two successive passages of the planet through the perihelion). The value of the factor $s$ for a perfectly elliptical orbit is unity, but we are anticipating the possibility that the function $u$ undergoes the dilation (stretching or shrinking) $u(\theta) \to u(s\theta)$ as a result of a perturbation.

With Eq.~(10), we have carried out a measurement. The operator $\hat{P}$ is tuned on the circular orbit $u = \kappa$ of the imaginary mean planet associated with the elliptical solution (7), and any variation of the common angular period of these planets—in the case of perturbed motion—will be detected.

The relative increment $\Delta \chi$ of the angular position of the perihelion over a complete orbit is obtained when we subtract from $\hat{P} u(s\theta)$ a full turn $2\pi$:
\begin{align}
\Delta \chi \equiv \hat{P} u(s\theta) - 2\pi = 2\pi \left(\frac{1}{s} - 1\right)\!.
\end{align}
Thus, the perihelion shift will be positive or negative according to whether $1/s$ is greater than or less than one.

Now let us multiply Eq.~(10) by a positive integer $n$. The effect of $n\hat{P}$ is equivalent to when the operation $\hat{P}$ is carried out $n$ times successively, assuming, for $1 \leq i \leq n$, that the terminal condition of the $(i - 1)$st operation becomes the initial condition for the $i$th operation:
\begin{align}\nonumber
n\hat{P} u &= \frac{n}{\kappa} \int_{\chi}^{\chi + 2\pi/s} u \, d\theta = \frac{1}{\kappa} \sum_{i=1}^{n} \int_{\chi + 2(i-1)\pi/s}^{\chi + 2i\pi/s} u \, d\theta\\& = \frac{1}{\kappa} \int_{\chi}^{\chi + 2n\pi/s} u \, d\theta \equiv \hat{P}^n u = 2\pi \frac{n}{s}.
\end{align}
Therefore, $n$ can be embedded in the upper limit of the integral, which we denote by the symbol $\hat{P}^n$, representing the $n$-fold composition of $\hat{P}$ with itself. Thus, the equality $\hat{P}^n u = 2\pi(n/s)$ is a consequence of the fact that the successive perihelia are evenly spaced with separation $2\pi(1/s)$.

\section{The Rotating Ellipse}
An ellipse in polar form is specified by the elements $a$, $e$, and $\chi$, which fix, respectively, the size, shape, and orientation of the ellipse in the plane. When a small perturbing force acts on the system, a viable approach is to treat these "constants" as variable. In our context, we should allow for variations of $\kappa$ and $\chi$, the elements $a$ and $e$ being contained in $\kappa$ (where they can vary independently).

In a specific application of this technique, the solution in the form $u(\theta) = \kappa + \kappa e \cos(\theta - \chi)$ is retained with $\chi$ no longer constant, but treated as a slowly varying function compared to $\cos \theta$. The function is a solution to a first-order differential equation solvable by successive approximations. For each value of $\chi(\theta)$, we can define an orbit, the osculating ellipse, with an orientation that varies according to the specific form of the function. In general, this is such that the perihelion is both oscillating and circulating, the latter eventuality being due to a part linear in $\theta$, which controls the secular perihelion shift.

The linear part produces a variation of the orbital period by a dilation of $u(\theta)$. To see this, assume $\chi(\theta) = (1 - s)\theta + \chi$, where $s$ is a real number (in actual applications $|s|$ will be very nearly equal to unity), and $\chi = \chi(0)$ is the initial position of the perihelion. Then:
\begin{align}\nonumber
\kappa + \kappa e \cos[\theta - \chi(\theta)]& = \kappa + \kappa e \cos \{\theta - [(1 - s)\theta + \chi]\}\\& = \kappa + \kappa e \cos(s\theta - \chi) = u(s\theta),
\end{align}
where $u(s\theta)$ is solution to the equation:
\begin{align}
u'' + s^2 u = \kappa.
\end{align}
Equation (13) shows that the phase of the sinusoid $u$ shifts in $\theta$ at a constant rate, so that in the plane $(\theta, u)$ the graph of $u$ can be imagined as a cosine wave traveling smoothly in the positive or negative $\theta$-direction at the uniform rate $1 - s$.

With the available tools, we can analyze some interesting qualitative aspects of a rotating ellipse. For what follows, it is better to visualize the orbit as represented in the polar plane $(r, \theta)$, where the perihelia lie on the circle of radius $r = a(1 - e)$. If we assume $1/s = 1 + g$ for $0 < g \ll 1$, then after $n$ orbital turns the relative shift of the perihelion from its initial position will be:
\begin{align}
\Delta_n \chi \equiv \hat{P}^n u(s\theta) - 2n\pi = 2n\pi g.
\end{align}
Thus, the operation $\Delta_n \chi$ maps the orbit $u(s\theta)$ to its $n$th tangency point with the circle of the perihelia. This point is identified by the angle $2n\pi g$, reckoned from an initial position $\chi$. We wish to consider $\Delta_n \chi$ as a dynamical system, and study the behavior of its ``orbits," i.e., we want to understand the behavior of the sequences $\Delta_n \chi$, for $n = 1, \dots, \infty$. Think of $n = 1, 2, \dots$ (rotation number) as a sequence of times. Then one can assimilate $\Delta_n \chi$ to a stroboscope that flashes briefly at these times, showing the planet at its successive perihelia.

Two different situations can occur. If $g$ is a rational number, say $p/q$, then $\Delta_q \chi$ flashes $2p\pi$ times until the perihelion is back to the start. This means that the orbital path closes; the orbit is periodic and takes the form of a rosette. If $g$ is an irrational number, the orbital path never closes and in the long run fills more and more densely the annulus $a(1 - e) \leq r \leq a(1 + e)$, in the sense that the trajectory of the planet intersects every neighborhood on the annulus, no matter how small. More specifically, the sequence of points $\Delta_n \chi$ will densely cover the circle of the perihelia, but each particular perihelion will never again be attained. It follows that the map $\Delta_n \chi$ is quasi-periodic, as made explicit by the following recurrence theorem: in the long run, a planet comes, on the circle of the perihelia, an infinite number of times arbitrarily close to any position already occupied. To demonstrate this, we can use the continued fraction approximations:
\begin{align}
g = \cfrac{1}{s_1 + \cfrac{1}{s_2 + \cdots}}.
\end{align}
Truncating at each successive stage gives an infinite sequence of rational approximates (the convergents):
\begin{align}
g \approx \frac{1}{s_1}, \quad \frac{s_2}{1 + s_1 s_2}, \quad \dots \quad \frac{p_1}{q_1}, \quad \frac{p_2}{q_2}, \quad \dots \quad \frac{p_n}{q_n}, \quad \dots,
\end{align}
where the integers $p_n$ and $q_n$ are coprime—their only common factor is 1—and $q_n > q_{n-1}$. The convergents $p_n/q_n$ play a role analogous to that of the partial sums of an infinite series. It can be shown that for each $n$, the difference from $g$ is less than $1/q_n^2$:
\begin{align}
\left| g - \frac{p_n}{q_n} \right| < \frac{1}{q_n^2},
\end{align}
and that these are the best rational approximations there are, in the sense that no rational fraction with a denominator not exceeding the denominator of the convergent does better.

Then, multiplying Eq.~(18) by $2q_n \pi$, we get:
\begin{align}
|2q_n \pi g - 2\pi p_n| = |\Delta_{q_n} \chi - 2\pi p_n| < 2\pi / q_n.
\end{align}
Now, for a given tolerance, however small, we can find a positive integer $n_0$ such that the middle term of Eq.~(19) is smaller for all values of $n$ greater than or equal to $n_0$. So we get closer and closer to $2\pi p_n$. By imagining suitable rotations of the circle of the perihelia (changes of origin), this reasoning extends to the generality of the perihelia.

\section{A Small Change of Curvature}
Suppose now that the constant $\kappa$, this structural component of the polar equation, is altered somewhat by adding—in an algebraic sense—a small piece $\delta \kappa$, thus affecting the maximal curvature of the orbital ellipse. This means that, for a given eccentricity, we have changed the semi-major axis of the elliptical orbit and the orbital radius of the mean planet.

Then the solution $u$ will become:
\begin{align}
u = (\kappa + \delta \kappa) + (\kappa + \delta \kappa)e \cos(\theta - \chi),
\end{align}
and, acting on it with $\hat{P}$, we obtain:
\begin{align}
\hat{P} u = 2\pi \left(1 + \frac{\delta \kappa}{\kappa}\right)\!.
\end{align}
The shift $2\pi \delta \kappa / \kappa$ will be an increase or a decrease, depending on the algebraic sign of $\delta \kappa$.

We wonder whether it is possible to build a suitable increment $\delta \kappa$ that encapsulates the presence of a central perturbing force. In this way, we would obtain a phenomenological derivation of the perihelion precession.

Notice that, while Eq.~(13) is a two-way relationship between $\chi(\theta)$ and $u(s\theta)$, an analogous, direct link between $\chi(\theta)$ and $\delta \kappa / \kappa$ does not exist. Because Eqs.~(10) and (21) both express a perihelion shift, we can establish indirectly a relationship between $\chi(\theta)$ and $\delta \kappa / \kappa$ via the dilation factor $s$, by identifying the right-hand sides of Eqs.~(10) and (21); in this way, we get:
\begin{align}
2\pi \left(\frac{1}{s}\right) = 2\pi \left(1 + \frac{\delta \kappa}{\kappa}\right)\!.
\end{align}
Now multiply both sides by $\kappa$. Then the area under the graph of $u(s\theta)$ in the interval $[0, 2\pi/s]$ is made equal to that of the rectangle of width $2\pi$ and of height $\kappa + \delta \kappa$. We assume that this area is an invariant of the perturbed system, in the sense that if we imagine bringing the height of the rectangle to the value $\kappa$, which reflects the geometry of the actual system, its width will vary by a factor $1/s - 1$. This indicates that, for the dynamic system, the addition of a $\delta \kappa$ should be understood as a virtual producer of a dilation of the function $u$.

As long as $|\delta \kappa| \ll \kappa$, from Eq.~(22), we get, to first order in $\delta \kappa / \kappa$:
\begin{align}
s = 1 - \frac{\delta \kappa}{\kappa},
\end{align}
and, by Eq.~(15) with $n = 1$, we have the identification $\delta \kappa / \kappa = g$. So either $s$ or $\delta \kappa$ can be used for the determination of the perihelion shift, and if we can connect one of them to the physics of the problem, we will have also determined the other, and vice versa. Further, we have:
\begin{align}
\chi(\theta) = (1 - s)\theta + \chi = \frac{\delta \kappa}{\kappa} \theta + \chi,
\end{align}
which expresses the secular shift of the perihelion in terms of the virtual relative increment of curvature of the elliptical orbit at its extrema in the presence of the perturbation.

\section{Relativistic Precession}
In dealing with the nonlinear Eq.~(3), we can imagine that the orbital equation containing all information on the perihelion precession of a planet, say Mercury, takes the linear form:
\begin{align}
u'' + u = \kappa_m, \quad \kappa_m = \text{const}.
\end{align}
In light of the results of Section VI, we need an interpretation of the symbol $\kappa_m$. We must assume that $\kappa_m$ is not exactly equal to the constant $\lambda/l^2$, which applies only to the inverse-square force. So we split $\kappa_m$ into two unequal parts: a dominant part $\kappa = \lambda/l^2$, and a much smaller secondary part $\delta \kappa \equiv \kappa_m - \kappa$, which encodes information about the function $3\alpha u^2$. With this interpretation, Eq.~(25) should be considered as an auxiliary linear equation whose particular integral is what interests us. Then Mercury will have the incremental orbital shift:
\begin{align}
\Delta \chi = 2\pi \frac{\delta \kappa}{\kappa}.
\end{align}
To implement this result, we rewrite the relativistic orbit equation as:
\begin{align}
u'' + u = \kappa + \underbrace{3\alpha u^2}_{\delta \kappa},
\end{align}
where we still do not know from where $\delta \kappa$ can come about.

We only know that the dimensionality of $\delta \kappa$ must be the inverse of a length, so that $\delta \kappa / \kappa$ is dimensionless. We tentatively guess $\delta \kappa \equiv 3\alpha \kappa^2$, obtained by substituting $\kappa$ for $u$ in the last term of Eq.~(27), and then we attempt a perturbative approach in which we take as the first approximate solution just the circular orbit that we have associated with the elliptical orbit. This procedure works, because from:
\begin{align}
u'' + u = \kappa + 3\alpha \kappa^2,
\end{align}
and from Eq.~(26), we get the Einstein formula:
\begin{align}
\Delta \chi = 6\pi \alpha \kappa = \frac{6\pi \alpha}{a(1 - e^2)}.
\end{align}
This shows that the relativistic precession is proportional to the maximum curvature $\kappa = a/b^2$ of the elliptical orbit.

Consider, for example, the actual figures for the orbit of Mercury. From $a = 0.3871 \, \text{AU}$ (one Astronomical Unit is exactly $149,597,811 \, \text{km}$) and $e = 0.2056$, we obtain:
\begin{align}
\kappa = \frac{1}{a(1 - e^2)} \approx 2.6973.
\end{align}
Additionally, from $M \approx 1.989 \times 10^{30} \, \text{kg}$, we get:
\begin{align}
3\alpha = \frac{3GM}{c^2} \approx 4.4309 \, \text{km},
\end{align}
which corresponds approximately to $2.96187 \times 10^{-8} \, \text{AU}$. Then:
\begin{align}
\delta \kappa = 3\alpha \kappa^2 \approx 2.1549 \times 10^{-7},
\end{align}
so that Eq.~(28) takes the numerical form:
\begin{align}
u'' + u = 2.6973 + 2.1549 \times 10^{-7}.
\end{align}
Then we get:
\begin{align}
\Delta \chi = 2\pi \frac{2.1549 \times 10^{-7}}{2.6973} \approx 5.0197 \times 10^{-7} \, \text{rad},
\end{align}
corresponding to $0.1035 \, \text{arc sec}$ per revolution. Mercury revolves about the Sun $415.2$ times in a century, so we have:
\begin{align}
\Delta \chi_{\text{sec}} = 0.1035 \times 415.2 = 42.97 \, \text{arc sec}.
\end{align}

Astronomical data show that the total dynamic secular perihelion shift of Mercury is about $574.096 \pm 0.41 \, \text{arc sec}$ per century, of which $531.50 \pm 0.85 \, \text{arc sec}$ is accounted for by the disturbances of the other planets. This corresponds, in Eq.~(30), to an additional numerical constant—evidently of order $10^{-6}$—that we shall calculate below to a precision within the relative standard observational uncertainty.

\section{Central Perturbing Forces}
Our approach to relativistic precession, supported by a chain of heuristic arguments, can be made systematic and general. Consider a polar orbital equation with a nonlinear term $\epsilon g(u)$, where $\epsilon$ is a parameter small enough to justify a perturbative approach. To this, we associate a linear equation with a constant term $\delta \kappa$, having the dimension of inverse length:
\begin{align}
u'' + u = \kappa + \underbrace{\epsilon g(u)}_{\delta \kappa}.
\end{align}

For the relativistic equation, where $\epsilon g(u) = 3\alpha u^2$, we have verified that the association $\epsilon g(\kappa) \to \delta \kappa$ works. However, such a simple recipe applies only to this case. Thus, we need a procedure that assigns to each specific function $\epsilon g(u)$ its $\delta \kappa$, while preserving the relativistic result.

To find this procedure, due to the dual role of $s$ and $\delta \kappa$, we first determine the dilation factor $s$ using a variational technique. Assume, in the presence of a perturbation, a small variation of the solution $u$ from the circular value $u = \kappa$, and then observe what happens. The variation equation can be formally obtained by applying the operator $\delta$ such that:
\begin{align}
\delta(u'') = (u + \delta u)'' - u'' = \delta u''.
\end{align}

Now consider Eq.~(44), where $g(u)$ is continuously differentiable in the closed interval $[u_{\text{min}}, u_{\text{max}}]$. Applying $\delta$ to both sides yields:
\begin{align}
\delta u'' + \delta u = \epsilon g'(u) \delta u.
\end{align}

Assuming the reference motion is the circular orbit $u = \kappa$, evaluate the derivative on the right-hand side at $\kappa$. This gives the homogeneous equation:
\begin{align}
\delta u'' + [1 - \epsilon g'(\kappa)] \delta u = 0,
\end{align}
which implies $s \approx 1 - \epsilon g'(\kappa)/2$. Thus, the lowest-order solution $u + \delta u$ to Eq.~(44) can be written in the form of the function $u(sh)$ of Eq.~(13), with $s = 1 - g$ and $g = \epsilon g'(\kappa)/2$.

It is interesting to note that, according to Eq.~(46), it is possible to replace in Eq.~(1) the actual perturbing force $\epsilon g(u) l^2 u^2$ with the inverse-cube force $\epsilon g'(\kappa) l^2 u^3$, in agreement with Newton’s theorem on revolving orbits \cite{deliseo2007a}. In conclusion, from Eqs.~(23) and (46), we get:
\begin{align}
\delta \kappa = \frac{1}{2} \epsilon \kappa g'(\kappa),
\end{align}
which establishes the correct relationship between $\delta \kappa$ and the function $\epsilon g(u)$ in the orbital equation. The resulting displacement of the perihelion per revolution will be given by:
\begin{align}
\Delta \chi = 2\pi \frac{\delta \kappa}{\kappa} = \pi \epsilon g'(\kappa).
\end{align}

This formula highlights the key role played by the maximum curvature of the ellipse in planetary precession. It also explains why substituting $u \to \kappa$ in $g(u) = u^2$ works only for the relativistic orbital equation. The reason is that if we assemble the initial-value problem:
\begin{align}
\frac{1}{2} \kappa g'(\kappa) = g(\kappa), \quad g(1) = 1,
\end{align}
we find that it has the unique solution $g(\kappa) = \kappa^2$, explaining why the relativistic perturbing force is the only one for which both approaches yield the same result.

The presence of the normalization factor $1/\kappa$ in the structure of the operator $\hat{P}$ suggests the formal simplification:
\begin{align}
\hat{P} u = \frac{1}{\kappa} \int u \, d\theta = \int \frac{u}{\kappa} \, d\theta \equiv \int v \, d\theta \equiv \hat{Q} v,
\end{align}
allowing us to write the orbital equations in dimensionless form, for which the circular mean orbit is $v = 1$. This device sometimes simplifies the mathematics and is often used for theoretical analysis. Dimensionality can be restored at any stage by reintroducing the factor $\kappa$.

The dimensionless form of Eq.~(44) is obtained via the substitution $u \to v$, $\kappa \to 1$, and then dropping $\kappa$ from $\delta \kappa$:
\begin{align}
v'' + v = 1 + \delta, \quad \delta \equiv \frac{1}{2} \epsilon g'(1).
\end{align}

Thus, employing the operator $\hat{Q}$, we have:
\begin{align}
\Delta \chi = 2\pi \delta.
\end{align}

To illustrate the use of this form in a simple application, we derive the perihelion shift arising by supposing that the exponent in Newton's law $l u^2$ is changed to a value slightly different from 2. This was one of the many pre-relativistic attempts to modify the law of gravitation to explain Mercury's motion \cite{hall1894}. Let us put $f(v) = l v^{2+\epsilon} = l v^2 v^\epsilon$ in the dimensionless form of Eq.~(1). If $\epsilon$ is small enough, we can limit ourselves to the first-order approximation:
\begin{align}
v^\epsilon \approx 1 + \epsilon \ln(v).
\end{align}

The resulting orbit equation becomes:
\begin{align}
v'' + v = 1 + \epsilon \ln(v).
\end{align}

By Eq.~(50), with $g(v) = \ln(v)$, we associate the equation:
\begin{align}
v'' + v = 1 + \frac{\epsilon}{2},
\end{align}
and so we get:
\begin{align}
\Delta \chi = \pi \epsilon.
\end{align}
Here, we do not need to make any dimensional adjustment, because $\epsilon$ is a pure number whose choice is made to fit Mercury's motion. To this lowest degree of approximation, the shift is the same for all planets.

\section{Newtonian Precession}

\subsection{The Model}
Now consider the perihelion precession caused by the gravitational pull of the other planets on Mercury. Strictly speaking, its exact determination involves the treatment of a three-dimensional many-body problem, while our perturbation approach is effective only for plane motions and central forces.

This difficulty can be overcome if we exploit the actual features of the planetary orbits—they are nearly coplanar and nearly circular—leading to rather realistic first approximations. Thus, on one hand, we can assume a common orbital plane. On the other hand, we will show that the cumulative effect of the forces exerted by each planet along its orbit on Mercury can be equated to that of a force of central type. It follows that we can use our tools with some minor, but clever, adaptations. Thus, we shall compute the precession of Mercury using an oversimplified, but surprisingly effective Copernican model, in which we assume the orbits of the other seven planets (from Venus to Neptune) to be circular and suitably spaced. In these circumstances, the average perturbing forces, those that interest us, are of central type and directed outward—conditions we know how to handle.

Under this assumption, we can write Mercury’s orbit equation in the form:
\begin{align}
u'' + u = \kappa + \sum_{n=0}^{7} \epsilon_n g_n(u),
\end{align}
where in the sum the index $n = 0$ applies to the relativistic term, while the remaining $n$ values apply to the effects of the other seven planets. We defer to the Appendix the calculations required for the determination of the function $\epsilon_n g_n(u)$ for a generic planet $n$. The conclusion is that Eq.~(44) takes the explicit nonlinear form:
\begin{align}
u'' + u = \kappa + 3\alpha u^2 - \sum_{n=1}^{7} \frac{m_n}{\kappa^2 r_n u} \left( \frac{r_n^2 u^2 - 1}{2} \right)\!,
\end{align}
where $r_n$ is the (constant) orbital radius of the planet $n$, and $m_n$ is the ratio of the planet's mass to the Sun's mass.

Thus, the small coupling constants of Eq.~(44) are $\epsilon_0 = 3\alpha$ and $\epsilon_n = m_n$ for $n = 1, 2, \dots, 7$. Because of the nonlinearity of Eq.~(45), we can resort to the principle of superposition of small disturbances, which is valid in the first-order mathematical treatment of the solar system. In our case, it can be stated by saying that the secular effect on the perihelion produced by the perturbing terms present in Eq.~(45) is the algebraic sum of those produced by each term taken singularly. The residue left out by this approximation is negligible. Therefore, our problem is to find the numerical form of the linear equation:
\begin{align}
u'' + u = \kappa + \sum_{n=0}^{7} \delta \kappa_n.
\end{align}
To do this for the planetary part, we first compute:
\begin{align}
m_n g_n'(u) = \frac{m_n}{\kappa^3 r_n^2 u^2} \left( \frac{3r_n^2 u^2 - 1}{2r_n u^2 (r_n^2 u^2 - 1)^2} \right)\!.
\end{align}
We must now specify the circular reference orbits to be included in the derivative.

\subsection{Effective Orbital Radii}
There is a sensible advantage in taking for Mercury not the orbit $u = \kappa$, the average with respect to the polar angle (which we used in the relativistic term), but the time average, which is $\bar{u} = 1/a$. On the other hand, the average distance, with respect to the angular variable $\theta$, of the planet $n$ from the Sun, is given by $\langle r_n \rangle = a_n$, the semi-major axis of its elliptical orbit. This follows from the definition of the ellipse, $r_n + d_n = 2a_n$, where $r_n$ and $d_n$ are the distances from the Sun and from the empty focus, respectively. The expression is symmetric in the two distances, so that their average values over an orbit are both equal to $a_n$. We shall use instead the time average:
\begin{align}
\bar{r}_n = a_n (1 + e_n^2 / 2).
\end{align}
This choice captures a dynamical aspect of the situation which would otherwise be excluded in a purely geometric treatment. As a consequence of the law of equal areas, the planet spends more time near the aphelion than near the perihelion. In an averaging process, the sample positions of the planet, per equal time intervals, are unevenly scattered over the elliptical orbit: they are grouped near the aphelion to a greater extent than near the perihelion, and therefore the average distance $a_n$ must be appropriately increased. It follows that if we want to approximate an elliptical orbit by a circle, we must use this effective radius.

\subsection{Perihelion Shifts}
When we insert the time average $\bar{u} = 1/a$ for Mercury and $\bar{r}_n$ for the planet $n$, Eq.~(47) becomes:
\begin{align}
m_n g_n'(\bar{u})|_{\bar{u} = 1/a} = \frac{m_n}{\kappa a^4} \left( \frac{3\bar{r}_n^2 - a^2}{2\bar{r}_n (a^2 - \bar{r}_n^2)^2} \right)\!.
\end{align}
Now we can make explicit the planetary portion of the last term of Eq.~(46) by using Eqs.~(36) (with $\kappa = 1/a$) and (48) to get:
\begin{align}
\sum_{n=1}^{7} \delta \kappa_n = \sum_{n=1}^{7} \frac{m_n}{\kappa a^3} \left( \frac{3\bar{r}_n^2 - a^2}{4\bar{r}_n (a^2 - \bar{r}_n^2)^2} \right)\!,
\end{align}
a rather tricky expression that summarizes a mess of mutual planetary positions.

We have carried out the calculation outlined in Eq.~(49) for each perturbing planet, and the results are presented in Table I. We have also displayed the constants associated with each planet, so its contribution can be verified. From the comparison with the results of more refined calculations—presented in the column labeled as "theory"—it is seen that our results are individually rather close to the correct ones.
\begin{table}[ht]
\centering
\small 
\setlength{\tabcolsep}{4pt} 
\resizebox{\columnwidth}{!}{
\begin{tabular}{|c|c|c|c|c|c|c|}
\hline
Planet & $\bar{r}_n$ & $m_n$ & $\delta \kappa_n$ & $\Delta \chi_n$ (arc sec) & Theory (arc sec) & Diff. (arc sec) \\
\hline
Venus & 0.7233 & 2.4478 & 1.3484003 & 268.72 & 277.85 & -9.13 \\
Earth & 1.0000 & 3.0404 & 0.4687968 & 93.44 & 90.04 & 3.40 \\
Mars & 1.5303 & 0.3227 & 0.0119692 & 2.35 & 2.53 & -0.18 \\
Jupiter & 5.2095 & 954.7786 & 0.7998580 & 159.44 & 153.58 & 5.86 \\
Saturn & 9.5511 & 285.8370 & 0.0386107 & 7.69 & 7.30 & 0.39 \\
Uranus & 19.2126 & 43.6624 & 0.0007229 & 0.14 & 0.14 & 0.00 \\
Neptune & 30.0701 & 51.8000 & 0.0002240 & 0.04 & 0.04 & 0.00 \\
\hline
Total & & & 2.6685819 & 531.82 & 531.48 & 0.34 \\
\hline
\end{tabular}
}
\caption{Mercury's secular perihelion shifts caused by the seven planets. The values of $m_n$ and $\delta \kappa_n$ are in units of $10^{-6}$.}
\label{tab:mercury_shifts}
\end{table}
The discrepancies in Table I should be mainly attributed to the fact that Eq.~(49) fails to take into account the non-central components of the perturbing forces. However, the differences are such that their algebraic sum is almost negligible: less than half a second of arc per century. We therefore make virtually no error if we use our total $\delta \kappa$ in writing the numerical form of the relativistic + Newtonian auxiliary equation (46) of the planet Mercury as:
\begin{align}
u'' + u = 2.6973 + 2.8841 \times 10^{-6}.
\end{align}
Thus, from one perihelion to the next, we have:
\begin{align}
\Delta \chi\! =\! \frac{2\pi}{\kappa} \sum_{n=0}^{7} \delta \kappa_n\! =\! \frac{2\pi (2.8841\! \times\! 10^{-6})}{2.6973}\! =\! 6.7183\! \times \!10^{-6} \, \text{rad},
\end{align}
corresponding to $1.3857 \, \text{arc sec}$ and to a centennial perihelion shift of $575.34 \, \text{arc sec}$. Our derivation yields an excellent fit to the observational data. Moreover, comparing the two numbers on the right-hand side of Eq.~(50) tells us the relative strength of the perturbing forces in comparison to the Sun’s inverse-square force: on the order of $10^{-6}$.

\subsection{A Proper Perspective}
In order to put our result into the proper perspective, it is worthwhile to analyze the effectiveness of our planetary model with circular orbits and time averages. A more realistic theory (still neglecting out-of-plane effects) could be done starting from the expansion in a Fourier series of the force (A5), where the tip of $r_n$ traces a Kepler ellipse. The coefficients of such a series, as is well known, are particular averages of the function to be expanded. If we confine ourselves to the secular effect on $\chi$ for the general case of a perturber moving in an elliptical orbit of eccentricity $e_n$, then in the first-order approximation we obtain an equation of the type:
\begin{align}
\chi'(\theta) = A + B e_n e \cos(\chi - \chi_n),
\end{align}
with two constant terms on the right side: a term that represents the average contribution of a nominal circular orbit, plus a correction term arising from the non-central part of the force, which takes into account the mutual orientations of the orbits of Mercury and planet $n$. The constants $A$ and $B$ depend on the semi-major axes of the two orbits, while $\chi$ and $\chi_n$ are the positions of the perihelia at a pre-fixed epoch. In the second term, a critical factor is the ratio of the eccentricities of the two planets. Because the eccentricity of Mercury is an order of magnitude greater than any other, one realizes that the contribution of the second term on the right-hand side of Eq.~(52) is small, which accounts for the minor corrective terms we found. Further, when considering the combined action of all planets, a random distribution of the perihelia attaches to each of these terms positive or negative signs, since the cosine runs through all its values from $-1$ to $+1$ as $\chi - \chi_n$ varies from $0$ to $2\pi$. Under opportune conditions, these terms nearly compensate when they are summed, as in the actual epoch. Because the planetary perihelia are all slowly moving, in another epoch the sum of the non-central terms of Eq.~(52) could produce a significant result of positive or negative sign, and our Copernican model would be less successful.

\section{Conclusion}
The method described here is essentially based on an averaging procedure, with all the advantages and limitations of an approach of this kind. It exploits the particular integral of a specific form of the orbital equation, to which is assigned a crucial role. By opportunely replacing the nonlinear term of the perturbed orbital equation with a constant, we build a virtual model of a fictitious planet on a circular orbit. The radius of this orbit differs very little—in a manner controlled by the perturbation—from the one in its absence. If we imagine traveling along the circular orbit of the unperturbed mean planet a distance equal to the circumference of the orbit of the fictitious mean planet, we will arrive slightly ahead of or behind the starting point. The operation $\Delta \chi$ extracts the angle, positive or negative, subtended by the small circular arc between the start and finish. The three worked examples have shown the validity of the method.

\appendix
\section*{Appendix: Average Force Exerted by a Planet on Mercury}
To avoid certain considerations about the center of mass, which do not affect the final result, we reduce the problem to the essentials. In the Sun-centered reference system, if Mercury is located at $\mathbf{r}$, the direct force per unit mass $\mathbf{f}_n$ exerted on it by planet $n$, located at $\mathbf{r}_n$, is:
\begin{align}
\mathbf{f}_n = \frac{\lambda_n (\mathbf{r}_n - \mathbf{r})}{|\mathbf{r}_n - \mathbf{r}|^3}, \quad r_n \equiv |\mathbf{r}_n| > |\mathbf{r}| \equiv r, \quad r_n = \text{const}.
\end{align}
Because we work in a plane environment, we can write the vectors in polar form via the complex exponentials:
\begin{align}
\mathbf{r}_n = r_n e^{i\theta_n}, \quad \mathbf{r} = r e^{i\theta}, \quad i = \sqrt{-1}.
\end{align}
If we define:
\begin{align}\nonumber
&c_n = \frac{r}{r_n} < 1, \quad \phi_n = \theta_n - \theta, \\& D_n = 1 + c_n^2 - 2c_n \cos\phi_n,\quad  d\phi_n = d\theta_n,
\end{align}
then Eq.~(A1) can be written in the form:
\begin{align}
\mathbf{f}_n = \frac{\lambda_n}{r_n^2} \left( e^{i\phi_n} - c_n \right) \frac{1}{D_n^{3/2}}.
\end{align}
The motions of the other planets and that of Mercury are rationally independent, in the sense that there is no simple numerical relationship between the periods. This means that the reciprocal positions on the respective orbits at any time are not related. It follows that we can consider the planet Mercury, at a generic position $\mathbf{r}$, to be affected by a secular force obtained by an averaging procedure.

At this point, it is useful to employ the factorization $D_n^{-3/2} = D_n^{-1} D_n^{-1/2}$. In fact, averaging with respect to $\phi_n$, the secular force at point $\mathbf{r}$ will be:
\begin{align}
\langle \mathbf{f}_n \rangle = \frac{\lambda_n}{2\pi r_n^2} \int_0^{2\pi} \frac{e^{i\phi_n} - c_n}{D_n} \, d\phi_n.
\end{align}
We now expand the function $D_n^{-1/2}$ in powers of $c_n$ and keep only the linear term:
\begin{align}
D_n(c_n)^{-1/2} \approx 1 + \frac{c_n^2}{2} (e^{i\phi_n} + e^{-i\phi_n}) + \cdots.
\end{align}
Then, after some algebra, Eq.~(A6) becomes:
\begin{align}
\langle \mathbf{f}_n \rangle = \frac{\lambda_n}{2\pi r_n^2} \int_0^{2\pi} \frac{e^{i\phi_n} - c_n/2}{D_n} \, d\phi_n.
\end{align}
Using the standard trigonometric integral:
\begin{align}
\int_0^{2\pi} \frac{e^{i\phi_n}}{D_n} \, d\phi_n = \frac{2\pi c_n}{1 - c_n^2}, \quad c_n < 1,
\end{align}
we finally get the average force exerted by the planet $n$ on Mercury:
\begin{align}
\langle \mathbf{f}_n \rangle = \frac{\lambda_n c_n}{r_n^2 (1 - c_n^2)} \hat{\mathbf{r}},
\end{align}
which is central and repulsive.

This function must be converted to a form $f_n(u)$ suitable for insertion in the orbital equation (1). To do this, we omit the unit vector $\hat{\mathbf{r}}$, then substitute $1/(r_n u)$ for $c_n$ and change the sign, obtaining:
\begin{align}
f_n(u) = -\frac{\lambda_n u}{2r_n (r_n^2 u^2 - 1)}.
\end{align}
It follows that the function to be inserted in the perturbation portion of Eq.~(1) is:
\begin{align}
\frac{f_n(u)}{l^2 u^2} \equiv \epsilon_n g_n(u) = -\frac{m_n}{\kappa^2 r_n u (r_n^2 u^2 - 1)},
\end{align}
where we have used $l^2 = \lambda/\kappa$, and so $m_n \equiv \lambda_n / \lambda$ is the planet-to-Sun mass ratio.


\begin{thebibliography}{99}

\bibitem{einstein1915} 
A. Einstein, Erklärung der Perihelbewegung des Merkur aus der allgemeinen Relativitätstheorie, Sitzungsberichte der Königlich Preußischen Akademie der Wissenschaften (Berlin, 1915), pp. 831–839; The Collected Papers of Albert Einstein, edited by A. J. Knox, M. J. Klein, and R. Schulmann (Princeton U. P., Princeton, NJ, 1996), Vol. 6, Doc. 24, pp. 112–116.

\bibitem{deliseo2011} 
M. M. D’Eliseo, “Higher-order corrections to the relativistic perihelion advance and the mass of binary pulsars,” Astrophys. Space Sci. 332, 121–128 (2011).

\bibitem{davies1983} 
B. Davies, “Elementary theory of perihelion precession,” Am. J. Phys. 51, 909–911 (1983).

\bibitem{gauthier1987} 
N. Gauthier, “Periastron precession in general relativity,” Am. J. Phys. 55, 85–86 (1987).

\bibitem{garavaglia1987} 
T. Garavaglia, “The Runge-Lenz vector and Einstein perihelion precession,” Am. J. Phys. 55, 164–165 (1987).

\bibitem{farina1987} 
C. Farina and M. Machado, “The Rutherford cross section and the perihelion shift of Mercury with the Runge-Lenz vector,” Am. J. Phys. 55, 921–923 (1987).

\bibitem{stump1988} 
D. R. Stump, “Precession of the perihelion of Mercury,” Am. J. Phys. 56, 1097–1098 (1988).

\bibitem{mcdonald1990} 
K. T. McDonald, “Right and wrong use of the Lenz vector for non-Newtonian potentials,” Am. J. Phys. 58, 540–542 (1990).

\bibitem{doggett1991} 
K. Doggett, “Comment on ‘Precession of the perihelion of Mercury,’ by Daniel R. Stump,” Am. J. Phys. 59, 851 (1991).

\bibitem{cornbleet1993} 
S. Cornbleet, “Elementary derivation of the advance of the perihelion of a planetary orbit,” Am. J. Phys. 61, 650–651 (1993).

\bibitem{dean1999} 
B. Dean, “Phase-plane analysis of perihelion precession and Schwarzschild orbital dynamics,” Am. J. Phys. 67, 78–86 (1999).

\bibitem{brill1999} 
D. R. Brill and D. Goel, “Light bending and perihelion precession: A unified approach,” Am. J. Phys. 67, 316–319 (1999).

\bibitem{deliseo2007a} 
M. M. D’Eliseo, “The first-order orbital equation,” Am. J. Phys. 75, 352–355 (2007).

\bibitem{lemmon2009} 
T. J. Lemmon and A. R. Mondragon, “Alternative derivation of the relativistic contribution to perihelic precession,” Am. J. Phys. 77, 890–893 (2009).

\bibitem{grossmann1996} 
N. Grossmann, The Sheer Joy of Celestial Mechanics (Birkhäuser, Boston, MA, 1996), p. 32.

\bibitem{dinverno2001} 
R. d’Inverno, Introducing Einstein’s Relativity (Oxford U. P., New York, NY, 2001), p. 194.

\bibitem{whiteside2008} 
The Mathematical Papers of Isaac Newton, edited by D. T. Whiteside (Cambridge U. P., Cambridge, 2008), Vol. III, 1670–1673, 169–173.

\bibitem{kendig2005} 
K. Kendig, Conics (The Mathematical Association of America, Washington, DC, 2005), p. 243.

\bibitem{szebehely1998} 
V. G. Szebehely and H. Mark, Adventures in Celestial Mechanics, 2nd ed. (Wiley, Hoboken, NJ, 1998), Chap. 11, pp. 221–245.

\bibitem{deliseo2012} 
M. M. D’Eliseo, “The quasi-elliptic motion of the Moon,” Chin. J. Phys. 50, 720–731 (2012).

\bibitem{olds1963} 
C. D. Olds, Continued Fractions (Random House, New York, NY, 1963).

\bibitem{clemence1947} 
G. Clemence, “The relativity effects in planetary motions,” Rev. Mod. Phys. 19, 361–364 (1947).

\bibitem{deliseo2007b} 
M. M. D’Eliseo, “Central forces and secular perihelion motion,” Can. J. Phys. 85, 1045–1054 (2007).

\bibitem{goldstein2002} 
H. Goldstein, C. Poole, and J. Safko, Classical Mechanics, 3rd ed. (Addison Wesley, San Francisco, CA, 2002), pp. 536–538.

\bibitem{hall1894} 
A. Hall, “A suggestion in the theory of Mercury,” Astron. J. XIV, 49–51 (1894).

\bibitem{roseveare1982} 
N. T. Roseveare, Mercury’s Perihelion from Le Verrier to Einstein (Clarendon Press, Oxford, 1982).

\bibitem{vandekamp1964} 
P. Van de Kamp, Elements of Astromechanics (Freeman, San Francisco, CA, 1964), p. 65.

\bibitem{murray1999} 
C. D. Murray and S. F. Dermott, Solar System Dynamics (Cambridge U. P., New York, NY, 1999), Appendix A, pp. 526–530.

\end{thebibliography}
\end{document}